\documentclass[aps,prl,twocolumn,letterpaper,superscriptaddress,10pt]{revtex4-1}
\usepackage{amsmath}
\usepackage{amssymb}
\usepackage{amsfonts}
\usepackage{txfonts}
\usepackage{bbold}
\usepackage{color}
\usepackage{bm}
\usepackage{graphicx}
\usepackage{enumerate}

\begin{document}
\title{Inverse Energy Cascade in Forced 2D Quantum  Turbulence} 
\author{Matthew T. Reeves}
\affiliation{Jack Dodd Centre for Quantum Technology, Department of Physics, University of Otago, Dunedin 9016, New Zealand}
\author{Thomas P. Billam}
\affiliation{Jack Dodd Centre for Quantum Technology, Department of Physics, University of Otago, Dunedin 9016, New Zealand}
\author{Brian P. Anderson}
\affiliation{College of Optical Sciences, University of Arizona, Tucson, AZ 85721, USA}
\author{Ashton S. Bradley}
\affiliation{Jack Dodd Centre for Quantum Technology, Department of Physics, University of Otago, Dunedin 9016, New Zealand}

\begin{abstract}
We demonstrate an inverse energy cascade in a minimal model of forced 2D quantum vortex turbulence. We simulate the Gross-Pitaevskii equation for a moving superfluid subject to forcing
by a stationary grid of obstacle potentials, and damping by a stationary thermal cloud. The forcing injects large amounts of vortex energy
into the system at the scale of a few healing lengths. A regime of forcing and damping is identified where vortex energy is efficiently transported to large length scales via an inverse energy cascade associated with the growth of clusters of same-circulation vortices, a Kolmogorov
scaling law in the kinetic energy spectrum over a substantial inertial range, and spectral condensation of
kinetic energy at the scale of the system size. Our results provide clear evidence that the inverse energy cascade phenomenon, previously observed in a diverse range of classical systems, can also occur in quantum fluids.
\end{abstract}
\maketitle

Turbulent fluid flows are characterized by the conserved transfer of kinetic
energy across length scales~\cite{Kolmogorov1941}.  In classical 3D turbulence, large-scale forcing
leads to a cascade of energy and vorticity towards smaller
scales~\cite{Sreenivasan1999a}.  In stark contrast, small-scale forcing in 2D
fluids may lead to an \textit{inverse} energy cascade
(IEC)~\cite{Kra1980.RPP5.547,Frisch1996}, associated with energy flux towards
larger scales, and the emergence of macroscopic rotating structures
from the turbulent flow. The IEC has been widely studied in classical
fluids~\cite{Sreenivasan1999a,Boffetta12a} and observed over a large range of
scales, from soap films~\cite{Martin1998a,Rutgers1998a} to planetary
atmospheres~\cite{Bouchet2002a}.  The IEC in classical fluids can be
studied via a point-vortex model in which the growth of large-scale flows
corresponds to the clustering of point-vortices with the same circulation~\cite{Eyink06a,SIGGIA1981a}. This model provides a link between
classical and quantum turbulence (QT), and suggests the intriguing possibility that the IEC may be observable in quantum
fluids. The
possibility to realize effectively 2D superfluid flows has motivated the study
of 2DQT in atomic Bose-Einstein condensates (BECs)~\cite{Parker05a, Nazarenko07a, Horng09a, Numasato10a,
Numasato10b, Sasaki2010,White10a, Nowak2011a, Nowak2012a, Schole2012a, Bradley2012a,
White2012a}, and such flows have been
experimentally demonstrated~\cite{Neely10a, Neely2012a}. Although intermittent
few-vortex clustering has been observed in the breakdown of superfluid
flow around an obstacle~\cite{Neely10a,Sasaki2010}, establishment of
a sustained, large-scale IEC in a BEC is hindered by vortex-antivortex annihilation, a consequence of the compressibility of the
superfluid~\cite{Numasato10a,
Numasato10b}. Additionally, compressible superfluids support a direct cascade of acoustic energy via weak-wave
turbulence~\cite{Nazarenko07a}. 
\par
In this Letter we demonstrate an IEC of quantum vortices in numerical
simulations of a forced, damped BEC.  We show that
the interplay between the effects of forcing and damping leads to a regime where a clear IEC occurs. For suitable damping acoustic energy is strongly suppressed, leading to approximately incompressible vortex dynamics analogous to the point-vortex model. We analyze the dynamics of the 2DQT via the
{\em incompressible} kinetic energy (IKE) spectrum~\cite{Nore97a,Numasato10b} associated with
vortices, and the statistics of vortex clustering as identified by established measures and a new
cluster-finding algorithm we develop. We observe three clear signatures of an IEC: (1) a
sustained increase in the charge and spatial scale of vortex clusters; (2) a Kolmogorov
$k^{-5/3}$ IKE spectrum~\cite{Nore97a} spanning an
inertial range of at least one decade; and (3) spectral condensation~\cite{Chertkov2007a} -- a macroscopic
accumulation of IKE at the largest length scales of
the system.

%============================================================================
\begin{figure*}
\begin{center}
\includegraphics[width=\textwidth]{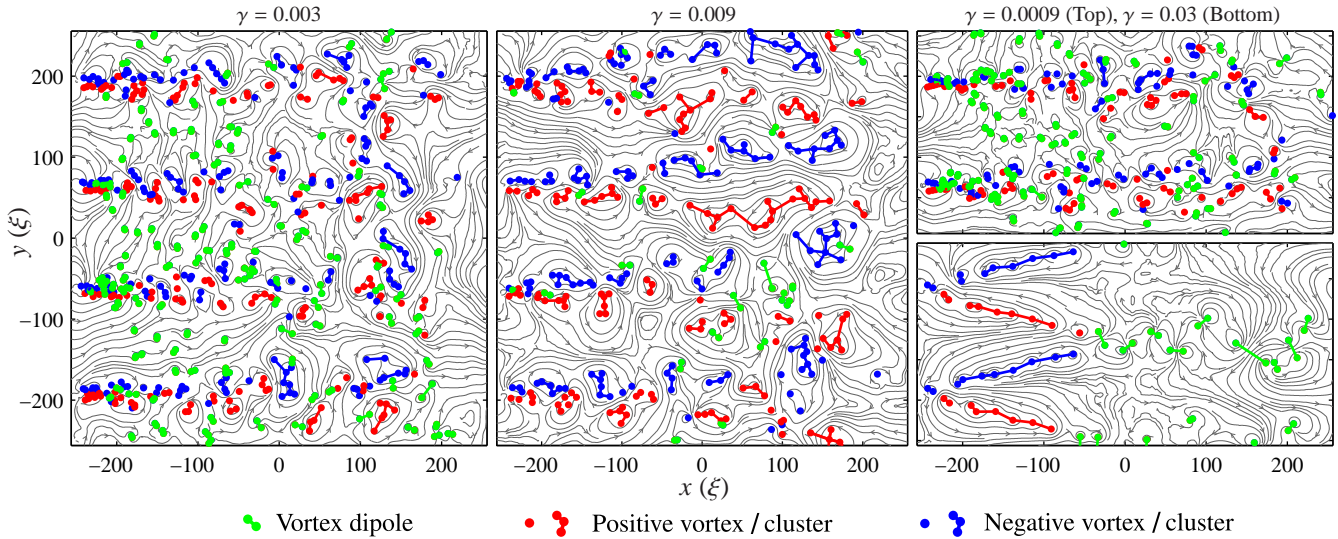}
\caption{\label{fig1}
(Color online) Vortex configurations in a rightward-flowing superfluid stirred by a stationary grid of obstacle potentials, located at the left. Dots indicate the locations of positive (counter-clockwise circulation) and negative (clockwise circulation) vortices at time $t=700\xi/c$. The vortices have been sorted into dipoles, free
vortices, and clusters (see legend) using the cluster-finding algorithm (see text). Vortices in dipoles at
the smallest length scales are not resolvable on the scale shown, and appear as
isolated green dots. Streamlines (grey
lines) give a visualization of the vortex-only velocity field by constructing the velocity field of an identical configuration of point vortices. The two right hand panels only show half of the computational domain.}
\end{center}
\end{figure*}
%============================================================================

Our system constitutes a BEC analog of forced 2D {\em grid turbulence}, a scenario
that has been used to demonstrate the IEC in classical
fluids~\cite{Rutgers1998a}.  We consider a superfluid moving with initial speed
$\mathrm{v}$ that is stirred by a stationary, regularly-spaced grid of
four narrow Gaussian potentials.  We model the dynamics of the condensate,
described by wavefunction $\psi(\mathbf{r},t)$, using the damped
Gross-Pitaevskii equation (dGPE)~\cite{Tsubota2002,*Penckwitt2002,*Blakie08a}
\begin{equation}
i\hbar \frac{\partial \psi (\mathbf{r},t)}{\partial t} = (1-i\gamma)(\mathcal{L} - \mu)\psi(\mathbf{r},t).
\end{equation}
 The operator $\mathcal{L}$ generates the zero-temperature GPE evolution, and
for a quasi-2D BEC subject to tight harmonic confinement in the $z$-direction
(with harmonic oscillator length $l_z$) can be written as
\begin{equation}
\mathcal{L} = -\frac{\hbar^2\nabla_\perp^2}{2m} + V(x,y) + g_2|\psi(x,y,t)|^2,
\end{equation}
where $V(x,y)$ is the grid potential, and $g_2 = 2 \sqrt{2\pi}\hbar^2
a_s/ml_z$, for atoms of mass $m$ interacting with $s$-wave scattering length
$a_s$. Dissipative collisions between condensate atoms and a stationary thermal cloud at
chemical potential $\mu$ are described by the dimensionless damping rate
$\gamma$. Imposing
periodic boundary conditions, our system provides a model of the
obstacle-induced spin-down of a persistent current immersed in a stationary
thermal cloud~\cite{Ramanathan11a}. 

We work in energy, length and time units of chemical potential $\mu$, healing
length $\xi = \hbar/mc$, and $\xi/c$ respectively, where $c=\sqrt{\mu/m}$ is
the speed of sound in the homogeneous system.  We numerically solve the dGPE in
a periodic box of length $L=512\xi$, for time $t=800\xi/c$ (at which time the
turbulent wake reaches $x=L/2$).  To eliminate the transient effects of abruptly raising the grid potential we
begin from a ground state of the dGPE where the grid is co-moving with the
flow, and stop the grid motion at $t=0$. The grid potentials are then located at $x=-L/2 + 8\xi$ and equally spaced by $L/4$ in
the $y$ direction.
We add a small amount of initial noise to
break the symmetry, and collect statistical information about the flow by
computing multiple trajectories.  Typical vortex configurations of the system
at time $t=700\xi/c$ are shown in Fig.~\ref{fig1}. To
observe turbulent flow, we choose sufficiently high ${\rm v} \approx 0.822c$
such that the grid nucleates vortices in a chaotic manner~\cite{Sasaki2010},
rather than periodically~\cite{Frisch92a}. To optimize
vortex injection, we choose grid potential height $V_0 = 100\mu$ and $1/e^2$
radius $w_0=\sqrt{8}\xi$. Each obstacle potential injects streams of opposite-signed vortices separated by distance $\sim 2w_0$. These streams carry negative momentum, arresting the superflow, and decay into clusters of charge $\sim 2$, such that the driving operates at an energy slightly higher than the Benard-von Karman transition point to turbulent injection~\cite{Sasaki2010}.

The dimensionless damping rate $\gamma$ influences the vortex dynamics,
and can be controlled via the temperature. 
In our forced system the stirring injects streams of small clusters of like-sign vortices that, together with damping, are vital for generating an IEC. At low damping (Fig.~\ref{fig1}, upper right panel), the vortex dynamics are dominated by mixing of these streams and dipole formation. In the over-damped regime (Fig.~\ref{fig1}, lower right panel) rapid separation of the vortex streams efficiently transfers linear momentum, spinning down the superflow, and suppressing turbulent dynamics. For intermediate damping (Fig.~\ref{fig1}, left and center panels) the streams are sufficiently polarized to suppress dipole formation, enabling clusters to merge and expand, while remaining in a regime of turbulent vortex dynamics. Associated with this phenomenology, we find that the short-wavelength ($k\xi>0.2$) compressible (acoustic) kinetic energy for $\gamma=0.009$ is suppressed by an order of magnitude below that for $\gamma=0.003$. Increasing $\gamma$ further does not alter it significantly. Consequently, for $\gamma=0.009$, sound is suppressed, while the downstream turbulent evolution is not appreciably altered by damping, in contrast to the over-damped regime. 

The expansion and merging of vortex clusters are the key signatures of an IEC in 2DQT~\cite{Bradley2012a,White2012a}.
To
quantify the vortex clustering seen in Fig.~\ref{fig1} we have developed a cluster-finding algorithm that identifies the vortex clusters within a given
configuration.  This yields more detailed spatio-temporal
statistical information regarding vortex clustering than
scalar measures \cite{White2012a}.  The algorithm consists of two rules: (1)
\textit{Opposite-sign vortices that are mutual nearest-neighbors constitute a
dipole and are removed from the algorithm's consideration}.  (2)
\textit{Same-sign vortices that are closer to each other than either is to an
opposite-sign vortex are placed in the same cluster}.  Rule 1 (2) is
applied recursively to all vortices still under consideration until no more
vortices can be added to dipoles (clusters). Rule 1 is applied first, removing the short-range velocity fields of dipoles. The
algorithm yields a decomposition of the configuration into clusters, dipoles,
and free vortices that corresponds extremely well to the streamlines of the point-vortex velocity field, as seen in Fig.  \ref{fig1}.  

%============================================================================
\begin{figure}
\begin{center}
\includegraphics[width=\columnwidth]{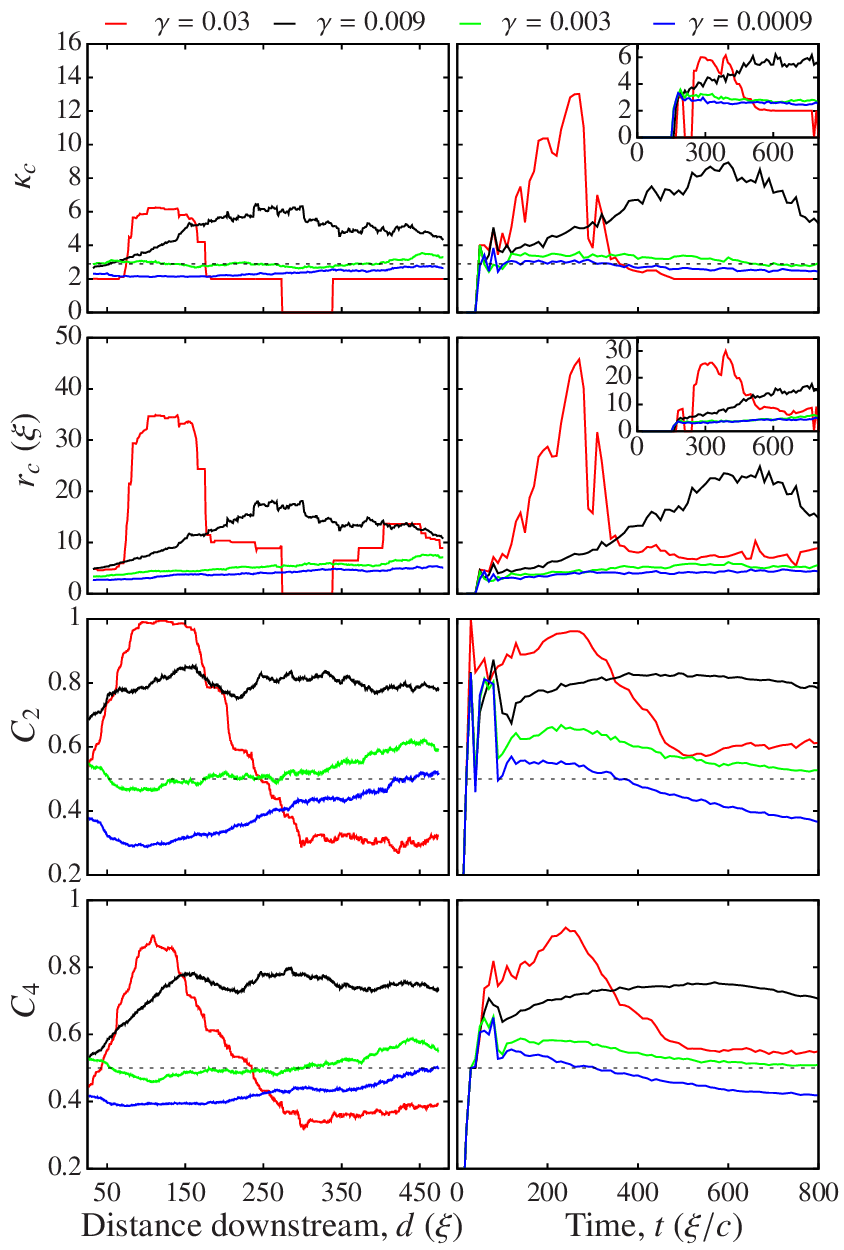}
\caption{\label{fig2}
(Color online) Spatio-temporal statistics of vortex clustering averaged over eight trajectories. For four values of $\gamma$ we
show the mean charge, $\kappa_c$, and radius, $r_c$, of clusters identified by
the cluster-finding algorithm and the 2nd and 4th-order nearest-neighbour correlation functions $C_2$ and $C_4$ defined in Ref.~\cite{White2012a}.
The left column shows clustering as a function of downstream distance from the grid, $d$, at time $t=700\xi/c$ obtained by counting only clusters with center in ($\kappa_c$ and $r_c$), or vortices lying in ($C_2$ and $C_4$), a box of
dimensions ($64\xi$,$512\xi$) centered at ($d-L/2$,$0$). 
The right column shows clustering as a function of time, $t$. For $\kappa_c$ and $r_c$, we
use a box centered at ($\mathrm{v}t-32\xi$,$0$) and hence co-moving with the flow. Insets
show the results for a delayed box
centered at ($\mathrm{v}t-96\xi$,$0$). For $C_2$ and $C_4$, we show the average value for the entire system.
Dashed horizontal lines at $\kappa_c = 2.9$ and $C_2=C_4=0.5$ indicate the expected value of these quantities for a random arrangement of vortices.}
\end{center}
\end{figure}
%============================================================================
 
In Fig.~\ref{fig2} we show how the mean charge (number of vortices),
$\kappa_c$, and radius (average distance of vortices from the cluster center),
$r_c$, of clusters identified by the algorithm, and the 2nd- and
4th-nearest-neighbor correlations $C_2$ and $C_4$ defined in
Ref.~\cite{White2012a}, change as a function of downstream distance from the
grid at time $t=700\xi/c$ (left column) and time (right column).  For $N$
vortices $C_B$ is equal to
$\sum_{i=1}^N \sum_{j=1}^B c_{ij}/BN$, where $c_{ij}=1$ if vortex $i$ and its $j$th nearest neighbor have the same sign, and $c_{ij}=0$ otherwise. The results of the spatial and temporal
analysis are in good qualitative agreement, consistent with the
frozen-turbulence approximation (equivalence of $d$ and ${\rm v}t$)~\cite{Frisch1996}. Our central result is shown by the strong
increases in all measures over large ranges of $d$ and $t$ in the case
$\gamma=0.009$, which are a clear signature of an IEC.  For lower values of the
damping ($\gamma=0.003$ and $\gamma=0.0009$) $\kappa_c$ remains generally
constant while $r_c$ increases slowly, providing no clear evidence for an IEC.
For higher damping ($\gamma=0.03$) the results are consistent with the
appearance of quasi-stationary linear clusters immediately behind the stirrers observed in Fig.~\ref{fig1}.
These clusters are analogous to stable (laminar) viscous shear layers in the
wake of the obstacles.  Simulations with larger $x$-domain ($L_x^\prime=2L$)
confirm that, for all values of the damping, the clustering statistics remain in the
same quasi-steady state for times up to the maximum propagation time
($L_x^\prime/{\rm v}$).

%============================================================================
\begin{figure}
\begin{center}
\includegraphics[width = \columnwidth]{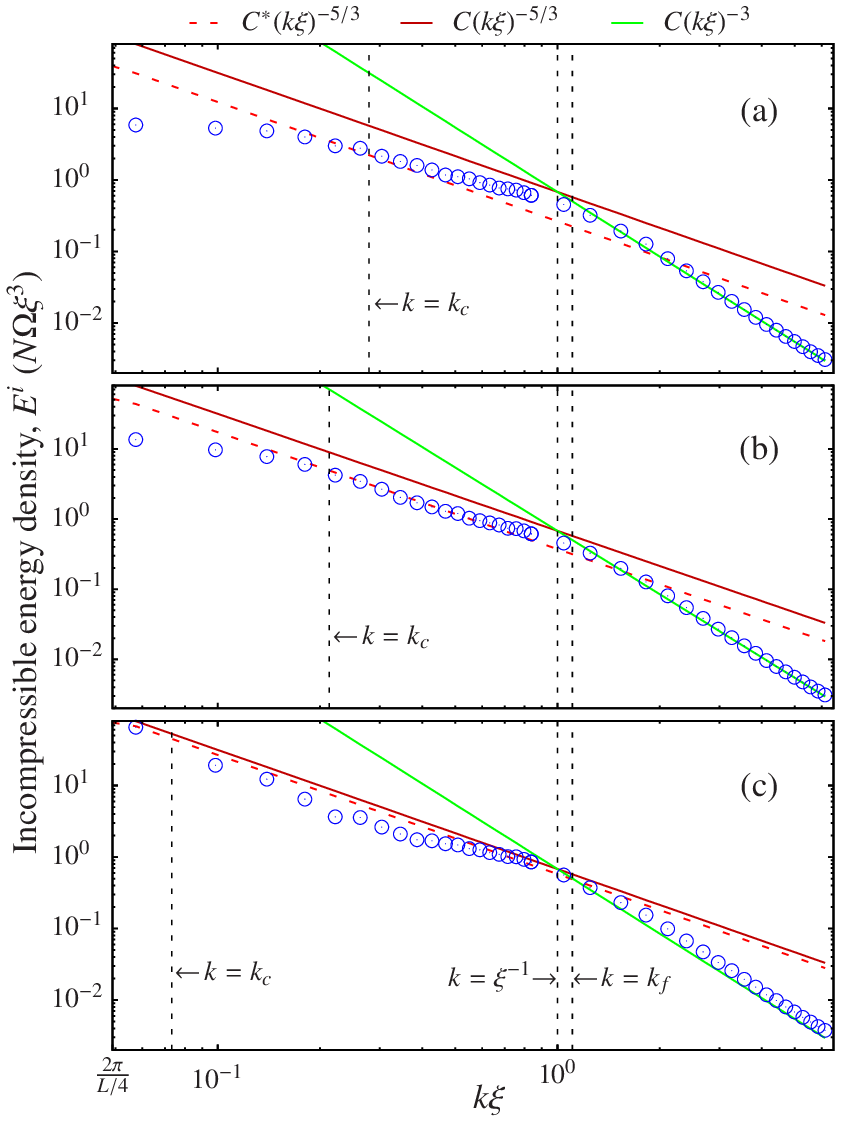}
\caption{(Color online) Incompressible kinetic energy spectra at $t = 550\xi/c$ for damping
parameters (a) $\gamma = 0.0009$ (b) $\gamma = 0.003$ (c) $\gamma = 0.009$
(circles), averaged over eight trajectories.  Lines show theoretical predictions for $k^{-3}$ [Eq.~(\ref{eq:EU})] and Kolmogorov $k^{-5/3}$ [Eq.~(\ref{eq:EC})]
spectra (see legend) in ideal 2DQT, with no fitted parameters. The Kolmogorov spectrum is shown using both the clustered
vortex number $N_c$ (constant $C^*$), and the
total vortex number $N$ (constant $C$) for comparison.
We also show wave numbers associated with the core size ($\xi^{-1}$), the forcing scale ($k_f$), and the mean diameter of the largest cluster ($k_c$). In an ideal IEC, $k_c$ approximately captures the lowest wavenumber of the Kolmogorov spectral region~\cite{Bradley2012a}.}
\label{fig3}
\end{center}
\end{figure}
%============================================================================

The IKE spectrum of our system (Fig.~\ref{fig3}) can be compared to the
spectrum for an {\em ideal} model of 2DQT given in Ref.~\cite{Bradley2012a}.
This model consists of quantum vortices with compressible cores;
incompressible energy is efficiently transported from small-scale forcing
towards large scales via vortex dynamics, but coupling of vortex dynamics to
the sound field is neglected.  In the ultraviolet (UV) region ($k\gg\xi^{-1}$)
the IKE spectrum is
\begin{equation}
E^i_{U}(k) = C(k\xi)^{-3} \equiv \Lambda^2 {N\Omega \xi^3} (k\xi)^{-3}, \label{eq:EU}
\end{equation}
where $\Lambda$ is a dimensionless constant related to the structure of a
compressible quantum vortex core, $N$ is the total vortex number, $\Omega =
2\pi\hbar^2n_0/m\xi^2$ is a constant with dimensions of enstrophy, and
$n_0=\mu/g_2$ is the homogenous system density.  The $k^{-3}$ form of
Eq.~(\ref{eq:EU}) is a universal feature of compressible 2DQT arising from the
finite size of a vortex core, and is not associated with a direct enstrophy
cascade. At the wave number $\bar{k}\equiv \xi^{-1}$ there
is a cross-over to point-vortex behavior in the infrared (IR) region
($L^{-1}\lesssim k\lesssim \bar{k}$).  Assuming (a) forcing wavenumber $k_f
\approx \bar{k}$, and (b) all $N$ vortices participate in clustering, the IR-region spectrum takes the Kolmogorov form
\begin{equation}
E^i_{C}(k) = C (k\xi)^{-5/3} \label{eq:EC}
\end{equation}
over an inertial range. The extent of the inertial range is linked to
the scale range of vortex clusters, and it has been shown that appropriately
distributed vortices in clusters of charge $\kappa_c=5$ are sufficient to
generate a decade of inertial range. 
In our simulations $k_f$ is determined by $w_0$: treating the injected vortices
as streams of dipoles of width $l_f=2w_0$, forcing occurs near $k_f=2\pi/l_f$,
corresponding to the peak in the IKE spectrum of such a vortex dipole~\cite{Bradley2012a}. Our
choice $w_0=\sqrt{8}\xi$ thus corresponds to $k_f=1.11\bar{k}$, satisfying
assumption (a).  Since the number of vortices participating in clustering, $N_c$, is generally less than $N$, assumption (b) is not always satisfied. However,
in this case the replacements $N\to N_c$ and $C\to C^*\equiv CN_c/N$ in
Eq.~(\ref{eq:EC}) give a better description of the inertial range, while
retaining continuity with the UV region at
$k\approx\bar{k}$.  
   
%============================================================================
\begin{figure}
\begin{center}
\includegraphics[width = \columnwidth]{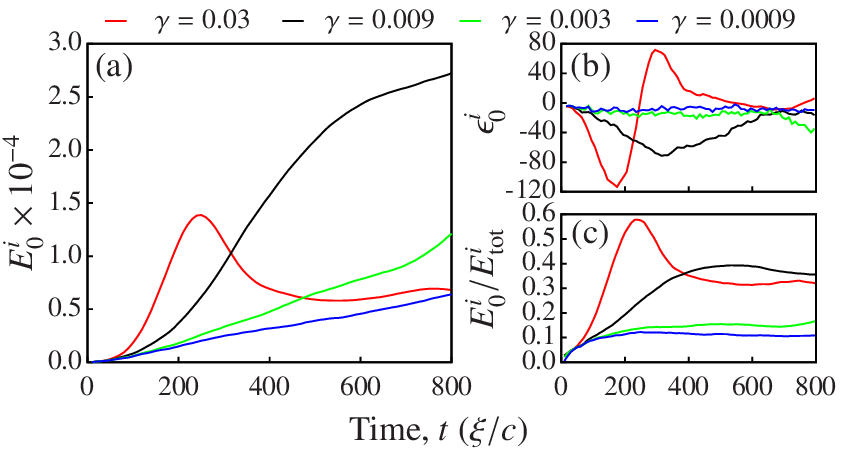}
\caption{(Color online) Signatures of spectral condensation. (a) Incompressible energy accumulated at the system scale ($E^i_0$, see text) and associated with the condensate. (b) Energy flux $\epsilon_0^i=-dE_0^i/dt$ from the condensate  to the rest of the system. (c) $E_0^i$ as a fraction of the total incompressible kinetic energy, $E^i_\mathrm{tot}=\int_{\Delta k}^\infty E^i(k) dk$. }
\label{fig4}
\end{center}
\end{figure}
%============================================================================

For all values of damping the UV spectrum agrees with the 2DQT prediction
[Eq.~(\ref{eq:EU})]. For $\gamma = 0.0009$ [Fig.~\ref{fig3}(a)] the IR spectrum
does not exhibit a clear $k^{-5/3}$ scaling and lacks temporal stability. A
buildup of energy around $\bar{k}$ is evident, suggesting a possible {\em
bottleneck}~\cite{Lvov2007a}, associated with inefficient energy transport away
from forcing wavenumber $k_f$.  For $\gamma = 0.003$ [Fig.~\ref{fig3}(b)] the
IR spectrum closely follows the modified 2DQT prediction [Eq.~(\ref{eq:EC})
with $N\to N_c$] over an inertial range of almost a decade of $k$, is steady
for times $300\xi/c \lesssim t \lesssim 600\xi/c$, and has a weaker bottleneck
around $k_f$. This is consistent with increased vortex clustering
(Fig.~\ref{fig1}), and suggests the onset of an IEC.  For $\gamma = 0.009$
[Fig.~\ref{fig3}(c)], the IR spectrum approximates the ideal 2DQT form
[Eq.~(\ref{eq:EC})] over a wide inertial range (at least one decade) and is
completely stable after $t \approx 400\xi/c$. This is consistent with strong
vortex clustering (Fig.~\ref{fig1}), and a fully developed IEC.  For strong
damping ($\gamma=0.03$) the spectrum has a large concentration of energy in the IR-region, but does not conform to the ideal 2DQT power-law.

Flux of IKE to scales approaching the system size leads to an effect known as \textit{spectral condensation}, 
associated with the emergence of macroscopic rotating structures~\cite{Chertkov2007a}. 
Previously, IKE fluxes in compressible, forced 2DQT have been calculated by
neglecting the compressible part of the kinetic energy, or introducing
assumptions regarding its coupling to the IKE spectrum \cite{Numasato10a,
Numasato10b}.  In Fig.~\ref{fig4} we provide a clear demonstration of a negative IKE flux by considering
the IKE accumulated at the system scale, given by $E^i_0 \equiv \int_{\Delta k}^{4\Delta k} E^i (k) dk$ where $\Delta k=2\pi/L$. Excluding the point at $k=0$ projects out any contribution from irrotational linear superflow; this is equivalent to calculating the spectrum in the rest frame of the fluid, and ensures that $E_0^i$ gives a robust signature of spectral condensation due to macroscopic vortex clustering.
In contrast to the other cases, for $\gamma=0.009$ we observe sustained rapid growth of the spectral condensate [Fig. 4(a)], corresponding to a consistent flux of energy towards the system scale [Fig. 4(b)], unambiguously demonstrating the IEC at this value of the damping.
In this case the turbulent IEC ultimately induces an even larger fractional condensation of energy than the very large laminar flow structures appearing in the over-damped case [Fig. 4(c)].

We have demonstrated an inverse energy cascade in a quantum analog of classical 2D grid turbulence~\cite{Rutgers1998a}. We solve the Gross-Pitaevskii equation describing a moving superfluid arrested by a grid of obstacle potentials, and damped by a stationary thermal cloud. We identify a regime of damping where quantum vortex clustering increases with time in the freely evolving superfluid turbulence downstream of the stirrers. This constitutes an inverse energy cascade, as corroborated by a Kolmogorov $k^{-5/3}$ incompressible kinetic energy spectrum,  and a flux of incompressible kinetic energy to large
scales. 
The demonstration of an inverse energy cascade of quantum vortices in a minimal model establishes a new link between classical and quantum turbulence, and 
our methods of analysis form a foundation for future studies of quantum vortex turbulence in two dimensional superfluids.
Our findings open new directions in the study of forced and decaying superfluid turbulence~\cite{Frisch92a,Sasaki2010} and provide a dynamical realization of the negative-temperature states for point vortices originally envisaged by Onsager~\cite{Eyi2006.RMP78.87}. 

\acknowledgments

This research was supported by the US National Science Foundation (PHY-1205713), The Marsden Fund of New Zealand (UOO162), and
The Royal Society of New Zealand (UOO004). 
\bibliographystyle{prsty}

\end{document}